\documentclass[11pt,a4paper]{article}
\usepackage[T1]{fontenc}
\usepackage[utf8]{inputenc}
\usepackage{lmodern,amsmath,amssymb,amsthm,mathrsfs,microtype,setspace,needspace}
\usepackage[margin=1.95cm]{geometry}
\usepackage{csquotes}
\usepackage[style=apa,natbib=true,backend=biber]{biblatex}
\usepackage{graphicx}
\usepackage{authblk}
\usepackage[hidelinks]{hyperref}
\newtheorem{proposition}{Proposition}
\newtheorem{corollary}{Corollary}
\newtheorem{assumption}{Assumption}
\newcommand{\dd}{\,\mathrm{d}}
\title{When Better Artificial Intelligence Widens the Adoption Gap in Health Care}
\author[1]{Jonas Fooken}
\author[1]{Ali Furkan Kalay\thanks{Corresponding author. Email: \texttt{alifurkan.kalay@mq.edu.au}}}
\author[2]{Metin Uyanik}
\affil[1]{Macquarie University Centre for the Health Economy}
\affil[2]{School of Economics, The University of Queensland}
\date{}
\begin{document}
\maketitle
\vspace{-1.5em}
\begin{abstract}
Artificial intelligence (AI) can assist healthcare providers in clinical practice or replace services for which they are paid. We model how provider influence shapes two policy choices. The first is adoption, the share of cases in which coverage is switched to care using AI. The second is the review requirement, the number of paid reviews that must accompany AI when it is used in clinical care. Assistance can lower providers' costs and increase their profits while leaving paid reviews unaffected. However, increasingly capable AI may support autonomous use that can replace reviews and reduce providers' profits. Provider gains from adoption first rise and subsequently fall as AI capability improves. Hence, providers oppose sufficiently autonomous AI if their lost income from reviews exceeds savings from better assistance. Opposition to AI adoption arises earlier for AI that displaces more reviews. When providers can influence adoption but not the review requirement, their opposition causes under-adoption. The patient value forgone at the margin grows as AI improves. At interior policies, provider influence over both decisions widens adoption and raises review requirements above the patient optimum.

\end{abstract}


\section{Introduction}
The adoption of artificial intelligence (AI) in health care reflects two institutional features. First, clinical expertise gives providers a role in coverage, payment, and regulatory decisions, making them partial gatekeepers to technologies that may assist or replace their work. Second, patients largely finance care collectively through taxes and insurance premiums. Payers make coverage and payment decisions on their behalf, so the costs are borne collectively rather than necessarily at the point of care. We model how provider influence shapes AI adoption and review requirements.

AI technologies are increasingly being adopted in health care. For example, US Medicare pays for an autonomous AI examination that issues a diagnostic report for diabetic retinopathy without physician review \citep{abramoffReimbursementFramework2022}. Yet in their 2020 comments to the US Food and Drug Administration (FDA), radiology's professional bodies supported assistive AI but opposed autonomous interpretation without physician confirmation, including systems that identify scans showing no abnormality and would allow them to bypass a radiologist's review \citep{acrRsnaFDA2020}.\footnote{Here and below, institutional examples illustrate where the model's margins arise. We do not attribute motives to particular decisions.} Under fee-for-service payment, this distinction matters because payment attaches to the professional's interpretation and written report \citep{cmsRepeatInterpretations2019}. Image reconstruction and automated measurements can lower the cost of an interpretation that the professional still performs and bills, whereas autonomous interpretation replaces the billed service. These effects give provider organisations a stake in adoption rules. Specialties closer to Medicare's fee-advisory committee obtain higher prices \citep{chanDicksteinIndustryInput2019}, and physician groups' political spending influences state scope-of-practice laws \citep{mcmichaelPoliticalSpending2017}.

Our main contribution is a task-based theoretical microfoundation for the change from support for AI adoption to opposition. Earlier models explain resistance through threatened political power or incumbent rents \citep{acemogluRobinsonPoliticalLosers2000,cominLobbiesTechnologyDiffusion2009}, and some find that resistance first rises and then falls with market size \citep{desmetParenteGuilds2014}. In our model, incumbents can initially be winners and subsequently losers. Using a task-based approach \citep{acemogluRestrepoAutomationNewTasks2019}, we derive the sign of providers' gain from the order in which AI reaches clinical tasks. This gain first rises with AI capability but falls when marginal savings per displaced review decline below the review margin. Opposition arises when eventual income losses from displaced reviews exceed retained savings. Provider influence therefore tilts adoption towards AI that assists paid work rather than replaces it, relative to its value to patients.

Following \citet{grossmanProtectionSale1994}, a policymaker weighs providers' lobbying contributions against patient value when setting adoption and review requirements. When a body insulated from provider influence sets reviews at the patient optimum, the influenced policymaker controls only adoption. Provider opposition then causes under-adoption, and the patient value forgone at the margin rises as AI improves. When the policymaker sets both decisions, at interior review policies, it requires reviews that patients would not buy. In the quadratic example, setting review requirements independently at the patient optimum reduces patient losses unless under-adoption is severe.

The results contribute to two strands of health economics. The first studies how reimbursement shapes providers' technology adoption \citep{bakerManagedCareMRI2001,selderPhysicianReimbursement2005,acemogluFinkelsteinInputTechnology2008,bakerLamiraudDRG2022,iversenMaTechnologyAdoption2022}, including second-best payment rules that, without lump-sum payments, accept overtreatment to speed adoption \citep{levaggiMorettoPertile2012}. We show that payment can also shape adoption rules through provider influence. The second studies physician supervision requirements. Relaxing restrictions on nurse practitioners has lowered prices \citep{kleinerRelaxingLicensing2016} and increased routine care \citep{traczynskiUdalovaNPIndependence2018}, while \citet{smithNPScopeOfPractice2022} finds no evidence of a change in patient volume. For nurse anaesthetists, \citet{dulisseCromwellNurseAnesthetists2010} find no evidence of increased inpatient deaths or complications following removal of the supervision requirement. Review requirements for AI are a new form of supervision requirement, and we show how provider influence can keep them above what patients would choose.

\section{Model}\label{sec:model}

A \emph{case} is one patient's service in which AI could be used, such as a chest X-ray. A \emph{review} is a professional's paid interpretation and written report, and \emph{providers} are the professionals paid for reviews and the practices that bill for them. \emph{Adoption} $a\in[0,1]$ is the share of cases, called \emph{adopted} cases, in which the payer switches coverage to care using AI, supplied by participating providers. Policy can set this share by indication, patient group, or care setting. For these cases, providers cannot instead bill for conventional care without AI. We assume that participation remains worthwhile at the practice level, so providers supply this care even when AI reduces income relative to existing care. Regulatory clearance is taken as given. The \emph{review requirement} $h\geq0$ is the expected number of reviews per adopted case, common to adopted cases because it is set for a type of use rather than case by case. We impose $h\leq1$, consistent with Medicare generally paying for only one interpretation of an emergency-room X-ray \citep{cmsRepeatInterpretations2019}. Thus $h$ can represent the share of AI-read X-rays requiring a radiologist's report. AI \emph{capability} $q\in\mathbb R$ indexes performance relative to typical task difficulty, so $q<0$ denotes capability below that benchmark. Information is complete and fees remain fixed as capability changes.

We evaluate policy from the perspective of (potential) patients, who largely fund care through taxes and premiums. \emph{Patient value} is the monetary value of health gains less spending on the costs of care, including review fees. Payments to providers count as costs, as in health technology assessment \citep{niceHTEManual2022}. Let $h^P(a,q)$ be the review requirement that maximises patient value at adoption $a$ and for capability $q$. With the number of cases normalised to one, the corresponding patient value relative to existing care, that is, care without AI, is $U(a,q)$. We assume $U(0,q)=0$ and $U_{aa}<0$, so wider adoption reaches cases where AI adds progressively less value.

\subsection{Provider gain: a task-based microfoundation}\label{sec:tasks}

Each review earns providers a margin $\theta>0$, the fee less the opportunity cost of the professional's time. The providers we model are paid per service. AI can first \emph{assist} with component tasks, such as image processing and measurement, lowering the provider's cost while retaining the paid report. Removing that report requires AI to perform the complete interpretation \emph{autonomously}, without professional review \citep{abramoffReimbursementFramework2022}.

Let $B(q)$ be the savings providers keep from AI-assisted tasks, such as preparing images, and $x(q)$ the paid reviews per case lost to autonomous AI. In the main model, the organised provider group retains these savings even after AI replaces its paid report, as an integrated practice operating the AI service might. This return comes from the AI service, not from billing for an unperformed review. The \emph{provider gain} per adopted case is $s(q)=B(q)-\theta x(q)$, the savings kept less the margin lost on displaced reviews. Providers support adoption when $s>0$ and oppose it when $s<0$. Expected provider gain is the same across adopted case types, although patient value and review needs may differ. The key object is the providers' saving per displaced review at the capability margin, $B'(q)/x'(q)$.

\begin{assumption}[Declining savings per displaced review]\label{ass:tasks}
$B$ and $x$ are continuously differentiable with $B'>0$ and $x'>0$, $B(-\infty)=x(-\infty)=0$, and finite limits $B(\infty)$ and $x(\infty)$. The ratio $B'/x'$ is strictly decreasing, with $\lim_{q\to-\infty}B'(q)/x'(q)>\theta>\lim_{q\to\infty}B'(q)/x'(q)$.
\end{assumption}

Proposition~\ref{prop:task} rests on this assumption. A common lag from assistance to autonomy and strictly log-concave task difficulty generate the declining ratio, which must also cross the review margin $\theta$ (Supplementary Section~S2.3). Assistance precedes autonomy because professionals can catch errors when AI assists, whereas autonomy requires accuracy high enough to dispense with review. In mammography, computer-aided detection has assisted radiologists since FDA approval in 1998 \citep{fentonCADMammography2007}, whereas AI-supported single reading for lower-risk mammograms was evaluated in the MASAI trial \citep{langMASAI2023}.

Consider reports whose difficulty $\xi$ is standard normal. AI can assist with a report's component tasks once $q\geq\xi$, saving $b>0$, but can replace the report only once $q\geq\xi+\delta$, with $\delta>0$. The expected number of paid reviews displaced per case approaches $e>0$ as capability rises. The distribution of difficulty is the same within every adopted case type, so the following are expected gains and displaced reviews per case. Savings persist after autonomy. Writing $\Phi$ for the standard normal distribution,
\begin{equation}
 B(q)=b\Phi(q),\qquad x(q)=e\Phi(q-\delta),\label{eq:paired}
\end{equation}
so $B'/x'$ falls from infinity to zero and Assumption~\ref{ass:tasks} holds for all positive $b$, $e$, $\delta$, and $\theta$ (Supplementary Section~S2.2), with $B(\infty)=b$ and $x(\infty)=e$. With one existing report per case, $e\leq1$. A single report with $e=1$ would give the same pattern in steps. The provider gain jumps to $b$ once AI reaches sufficient capability to assist and falls to $b-\theta e$ when it becomes autonomous.

Differences in difficulty across reports smooth these steps, and the normal shape places most reports at moderate difficulty. Supplementary Section~S2.4 derives the same rise and fall in provider gain when replacing a report requires several component tasks to be autonomous. Figure~\ref{fig:rollout} illustrates the example with $b=1/4$, $e=1$, $\delta=1$, and $\theta=1/2$.

\subsection{Political choice}

In our model, there is one organised provider group that offers contributions to the policymaker conditional on the policy chosen. The policymaker can be a public payer, such as Medicare, or a private health insurer, and can regulate both adoption and review rules. Contributions represent any form of influence, including official advisory roles, lobbying and public communications. Following \citet{grossmanProtectionSale1994}, the policymaker maximises total provider gain plus $\lambda$ times patient value, both summed over adopted cases:
\begin{equation}
 J(a,h;q)=a\,[s(q)+\theta t]+\lambda\Big[U(a,q)-\frac{c}{2}\,a\,t^2\Big],\qquad t\equiv h-h^P(a,q),\qquad \lambda,c>0.\label{eq:J}
\end{equation}
Providers earn $s$ for each adopted case plus the margin on \emph{excess reviews} $t$, the reviews required beyond the patient optimum $h^P$. The margin on reviews up to $h^P$ is already reflected in $s$. Reviews go first to the cases where they matter most, so patient value falls as reviews depart from $h^P$, at a rate set by $c$. Too few reviews forgo health gains worth more than their fees, and too many buy gains worth less. The policy that maximises $J$ is the one selected in the single-group menu-auction equilibrium \citep{bernheimMenuAuctions1986}. The weight $\lambda$ measures how the policymaker weighs benefits from providers and patients, giving up a dollar of patient value in return for at least $\lambda$ dollars of provider gain. As $\lambda$ grows, policy approaches the patient optimum, our benchmark.

\section{Results}\label{sec:results}

\subsection{Support, then opposition}

\begin{figure}[!t]
\centering
\includegraphics[width=\textwidth]{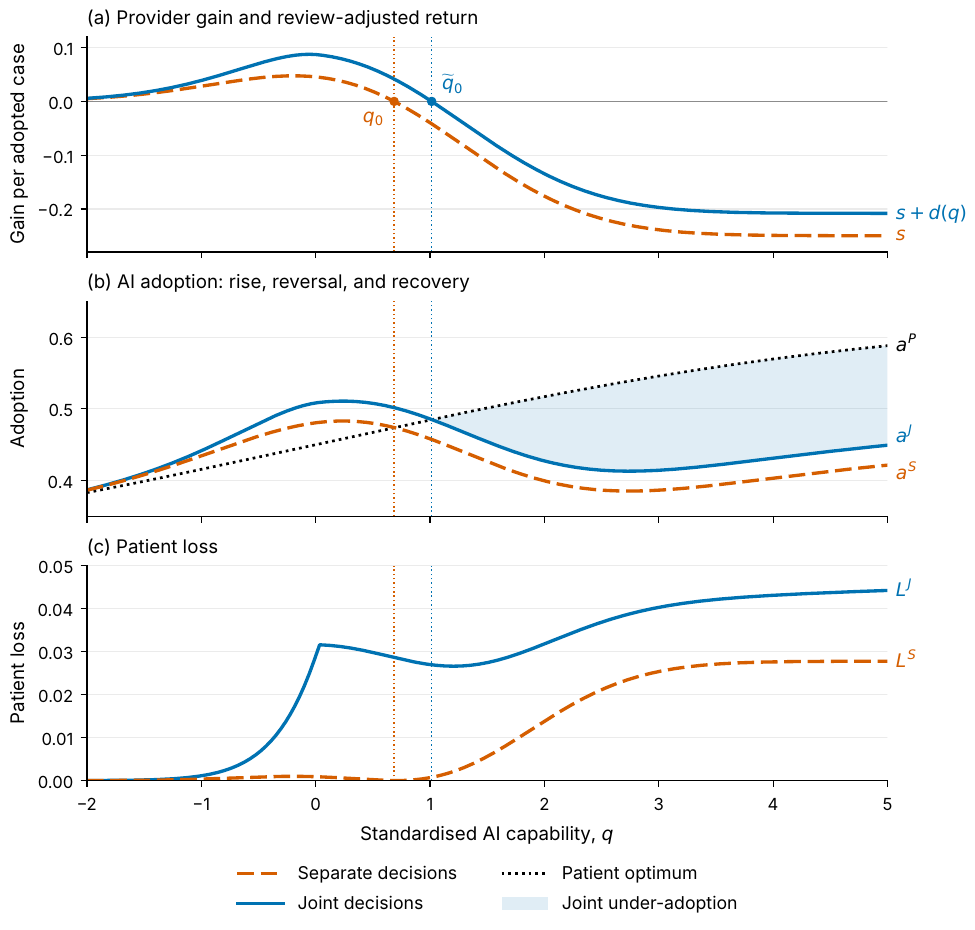}
\caption{Provider gain and review-adjusted return, AI adoption, and patient losses.}\label{fig:rollout}
\begin{minipage}{0.97\textwidth}
\footnotesize
\textit{Notes:} Example \eqref{eq:paired} with quadratic patient value and the parameters in Supplementary Section~S4.3. Capability is measured relative to mean task difficulty, so $q<0$ is below that benchmark. Dotted lines mark the onset of under-adoption under separate ($q_0$) and joint ($\widetilde q_0$) decisions. Panel (a) shows provider gain $s$ and the review-adjusted return $s+d(q)$, where $d(q)=\theta t^J(q)-\lambda c[t^J(q)]^2/2$. Provider gain under joint decisions is $s+\theta t^J$; the return deducts the weighted patient cost of excess reviews. The kink in panel (c) occurs where the one-report cap ceases to bind, at $q\simeq0.033$.
\end{minipage}
\end{figure}

The provider gain rises while savings on newly assisted tasks exceed the income lost from displaced reviews, $B'/x'>\theta$. Providers can save time preparing images or taking measurements while keeping paid interpretation. Once $B'/x'<\theta$, further capability improvements displace more review income, $\theta x'(q)$, than they add in savings, $B'(q)$, so total gain $s(q)$ falls although savings keep accumulating. Opposition requires eventual lost review income, $\theta x(\infty)$, to exceed full savings, $B(\infty)$ (in example \eqref{eq:paired}, $\theta e>b$). This is plausible in screening, where most examinations show no abnormality. Autonomous reporting could remove these reviews, whereas assistance only shortens them. Figure~\ref{fig:rollout}(a) shows the gain rising, peaking at $\widehat q$, and turning negative at $q_0$.

\begin{proposition}[Provider support]\label{prop:task}
Under Assumption~\ref{ass:tasks}, the provider gain $s=B-\theta x$ satisfies $s'(q)>0$ for $q<\widehat q$ and $s'(q)<0$ for $q>\widehat q$, where $B'(\widehat q)/x'(\widehat q)=\theta$. A finite zero $q_0$ exists if and only if $\theta x(\infty)>B(\infty)$. Then $q_0>\widehat q$ and $s(q)<0\iff q>q_0$.
\end{proposition}

AI technologies differ in these parameters, i.e., the savings $b$, the reviews displaced $e$, and the lag $\delta$ between assistance and autonomy. AI that takes measurements on an X-ray mainly saves time, a high $b$, whereas AI that reports normal X-rays removes reviews, a high $e$. Corollary~\ref{cor:statics} shows how these parameters move the opposition threshold $q_0$.

\begin{corollary}[Selection of AI technologies]\label{cor:statics}
In example \eqref{eq:paired}, if $\theta e>b$, the threshold $q_0$ rises with $b$ and $\delta$ and falls with $e$ and $\theta$.
\end{corollary}

Provider influence therefore affects which type of AI is adopted earlier or to a greater extent. Technologies that displace more reviews per report (higher $e$) meet opposition at lower capability $q$, whereas assistive technologies that generate large savings to providers (higher $b$) or have a long lag before autonomy (higher $\delta$) keep provider support to higher $q$. Holding patient value fixed, provider influence thus tilts adoption toward assistive AI.

\subsection{Under-adoption}\label{sec:separate}

Superscripts mark the decision, with $P$ for the patient optimum, $S$ for separate decisions, and $J$ for joint decisions. When adoption $a$ and reviews $h$ are regulated separately, the review requirement is set at the patient optimum $h^P(a,q)$ by a body that providers cannot influence, implying $t=0$. Patient-optimal adoption $a^P$ solves $U_a(a^P,q)=0$, whereas the policymaker chooses $a^S$ to maximise $\lambda U(a,q)+a\,s(q)$. The \emph{adoption gap} $U_a(a,q)$ is the patient value of the marginal adopted case. It is zero at $a^P$, positive under under-adoption, when extending AI to further cases would still benefit patients, and negative under over-adoption.

\begin{corollary}[Separate decisions]\label{cor:rollout}
At interior policies,
\begin{equation}
 U_a(a^S,q)=-\frac{s(q)}{\lambda}.\label{eq:separate}
\end{equation}
Hence $a^S<a^P$ if and only if $\theta x(\infty)>B(\infty)$ and $q>q_0$, and for such $q$ the adoption gap $U_a(a^S(q),q)$ widens as $q$ rises.
\end{corollary}

By Corollary~\ref{cor:rollout}, adoption stops where the patient value of the marginal case, weighted by $\lambda$, offsets the provider's loss. Once under-adoption begins, each improvement in capability $q$ displaces more review income than it adds in savings, so more capable AI widens the adoption gap. The gap is also larger for AI that displaces more reviews, as reflected by a higher $e$ in example \eqref{eq:paired}. When the provider gain is positive, as for $q<q_0$, providers push adoption beyond the patient optimum, leading to over-adoption, as in Figure~\ref{fig:rollout}(b).

\subsection{Excess reviews}\label{sec:joint}

When $a$ and $h$ are regulated jointly, the policymaker sets both adoption, $a^J$, and the review requirement, $h^J$ (Proposition~\ref{prop:policy}). Each excess review earns providers the margin $\theta$, at a rising cost to patients.

\Needspace{10\baselineskip}
\begin{proposition}[Joint decisions]\label{prop:policy}
At interior policies, excess reviews and adoption satisfy
\begin{align}
 t^J\equiv h^J(a,q)-h^P(a,q)&=\frac{\theta}{\lambda c},\label{eq:hI}\\
 U_a(a^J,q)&=-\frac{s(q)+\theta t^J/2}{\lambda}.\label{eq:joint}
\end{align}
Hence $a^J>a^S$, and $a^J<a^P$ if and only if $\theta x(\infty)-B(\infty)>\theta t^J/2$ and $q>\widetilde q_0$, where $\widetilde q_0>q_0$ solves $s(\widetilde q_0)=-\theta t^J/2$.
\end{proposition}

Excess reviews compensate providers for the review income they lose to AI. A policy may, for example, pay for AI to read an X-ray but still require a professional report, even where that report adds less health benefit than it costs. Patients bear this cost in every adopted case.

Excess reviews also increase adoption. At an interior review policy, each adopted case now brings the policymaker the provider gain plus $\theta t^J/2$. This is the margin on excess reviews, $\theta t^J$, less their weighted cost to patients, $\lambda c(t^J)^2/2=\theta t^J/2$. Under-adoption therefore begins only once the gain falls below $-\theta t^J/2$, and never if $\theta t^J/2$ exceeds providers' eventual loss from adoption, $\theta x(\infty)-B(\infty)$. In Figure~\ref{fig:rollout}(a), the review-adjusted return crosses zero later, at $\widetilde q_0$. 
In Figure~\ref{fig:rollout}, the cap $h\leq1$ affects this gain only at lower capability, where most cases still need a review and little room remains for excess reviews. More generally, the cap can remain binding at high capability; the interior adoption and threshold results then need not apply (Supplementary Section~S3.2).
Because $a^J>a^S$ at interior review policies, excess reviews also increase over-adoption while the provider gain is positive. Patients pay for this wider adoption, and their aggregate loss is
\begin{equation}
 L^J=U(a^P,q)-U(a^J,q)+\frac{c}{2}\,a^J(t^J)^2.\label{eq:loss}
\end{equation}
In \eqref{eq:loss}, $U(a^P,q)-U(a^J,q)$ is the loss from adoption differing from the patient optimum. The remaining term is the cost of excess reviews across adopted cases, which persists even at patient-optimal adoption.

To compare the two decisions, take an example in which patient value is quadratic in adoption, $U(a,q)=V(q)a-ka^2/2$ with $k>0$. At interior policies,
\begin{equation}
 a^P-a^S=-\frac{s(q)}{\lambda k},\qquad
 a^J=a^S+\frac{\theta t^J}{2\lambda k}.\label{eq:main-gap}
\end{equation}
The adoption gap is then $U_a(a,q)=k(a^P-a)$. Under separate decisions it rises whenever $s$ falls, and under joint decisions it does so when the cap $h\leq1$ does not bind. Adoption itself falls only if $-s'(q)>\lambda V'(q)$ in that region. Figure~\ref{fig:rollout}(b) shows adoption rising, falling, and recovering while the adoption gap keeps widening after the cap ceases to bind. Let $L^S=U(a^P,q)-U(a^S,q)$ be the patient loss under separate decisions. In this example, $L^S<L^J$ whenever $a^S\geq a^P/2$ because the extra adoption due to excess reviews is only valuable when the shortfall is large. Figure~\ref{fig:rollout}(c) shows separate decisions losing less at every capability $q$.

\section{Discussion}\label{sec:discussion}

Our results imply that review requirements for AI deserve the same scrutiny as adoption, because broad adoption can come bundled with reviews patients would not buy. How providers are paid matters through the margin $\theta$ on each review, which raises both the income lost on displaced reviews and the value to providers of excess reviews. Under a fixed payment per case, an additional review brings no additional fee but still costs the professional's time, so the margin $\theta$ is negative. Provider influence therefore pushes the review requirement below the patient optimum (Supplementary Section~S5.3), matching the skimping concern of prospective payment \citep{ellisMcGuireProspectiveReimbursement1986,ellisCreamingSkimping1998}. Compensating providers through excess reviews can cost patients less than paying them directly per AI use, and may be less visible to voters (Supplementary Section~S5.1). \citet{levaggiMorettoPertile2012} obtain a similar trade-off for a welfare-maximising payer without lump-sum payments, whereas here it arises from provider influence.

\subsection{Applications}\label{sec:applications}

Three applications illustrate the payment margins rather than identify provider motives. Computer-aided detection in mammography assists radiologists while retaining their interpretation. Its use spread after Medicare began paying for it in 2001 \citep{fentonCADDownstream2014}, although large studies found no gain in diagnostic accuracy \citep{fentonCADMammography2007,lehmanCADMammography2015}. Autonomous retinal screening can expand access beyond existing eye-care provision \citep{abramoffReimbursementFramework2022}. If it mainly reaches patients who would otherwise go unscreened, displaced review income, $\theta x$, is small. Autonomous reporting of normal chest X-rays was evaluated retrospectively in Danish public hospitals for its potential to reduce radiologist workload \citep{plesnerAutonomousCXR2023}. Hospital doctors there are salaried, with pay based on hours rather than clinical output \citep{observatoryDenmarkPayment2024}. Replacing a reading therefore does not directly remove a per-reading fee from the radiologist.

\subsection{Other explanations}\label{sec:other}

Other explanations also predict resistance to autonomous AI. Performance may be uncertain or vary across sites, as the radiology bodies argued \citep{acrRsnaFDA2020}. The precautionary value of review then belongs in $h^P$, and influence adds reviews beyond it. Radiologists may also have difficulties in combining their judgement with AI predictions \citep{agarwalHumanExpertiseAI2023}, and liability law tends to favour following the standard of care over following an AI recommendation \citep{priceGerkeCohenLiability2019}. None of these explanations links review requirements to how providers are paid, and our model provides that link.

\subsection{Separating professionals and billing organisations}\label{sec:separation}

We model an integrated provider. If hospitals or other organisations that bill for services keep the savings from greater AI adoption while only professionals are organised, opposition starts earlier, because the professionals lose review income without keeping the offsetting savings. A paid-report requirement then also becomes a more valuable concession to those professionals, because it restores both their review margin and the savings from assisted care (Supplementary Section~S5.2). The ranking in Figure~\ref{fig:rollout}(c) can then reverse, for about $2.85<q<4.12$ at the figure's parameters if professionals keep none of the return from AI-only care.

\section{Conclusion}

We make three qualitative predictions. First, for given capability $q$ and savings $B$, opposition to autonomous AI should be stronger when providers are paid per review than when they are salaried or paid per case. Second, under fee-for-service payment and at interior policies, review requirements attached to AI should exceed the patient optimum, by more where provider organisations have greater influence over the payer. Third, provider influence should favour assistive over autonomous AI relative to their value to patients, even when autonomous AI benefits patients. Overall, AI that first assists providers and later replaces their paid reviews can turn them from supporters into opponents of adoption. Requiring paid reviews alongside AI can keep their support for longer, but patients pay for reviews they would not buy. Better AI therefore may not bring policy closer to what patients would choose.

\section*{Data availability}

No new data were created or analysed in this study. The figure is computed directly from the parameter values reported in the Supplementary Material, and the code that generates it is provided with the Supplementary Material.

\clearpage
{\small\printbibliography}

\clearpage
\appendix
\renewcommand{\thesection}{S\arabic{section}}
\renewcommand{\theHsection}{S\arabic{section}}
\counterwithin{equation}{section}
\renewcommand{\theHequation}{S\arabic{section}.\arabic{equation}}
\setcounter{corollary}{0}
\renewcommand{\thecorollary}{S\arabic{corollary}}
\renewcommand{\theHcorollary}{S\arabic{corollary}}
\begin{center}
{\Large Supplementary Material}
\end{center}

Section~\ref{app:accounting} derives the aggregate model from payment accounting. Section~\ref{app:tasks} proves Proposition~\ref{prop:task} and Corollary~\ref{cor:statics}, and extends the example to bundled component tasks. Section~\ref{app:proofs} gives the political equilibrium and proves Corollary~\ref{cor:rollout} and Proposition~\ref{prop:policy}. Section~\ref{app:nesting} develops the quadratic example and Figure~\ref{fig:rollout}. Section~\ref{app:extensions} treats direct payments, ownership of the AI-only return, nonpositive margins, and case-specific provider gains. Numbers without an S prefix refer to the main text. All comparisons hold the specified primitives fixed.

\section{Foundations and aggregate representation}\label{app:accounting}

\subsection{Payment accounting}

Coverage is switched to care using AI for adopted cases, which participating providers supply. Conventional care without AI is not reimbursable for those cases. Participation constraints at the practice level, including the cost of contributions, do not bind for the policies considered, so a negative provider gain below is a loss relative to existing care, not an exit. The integrated practice continues to supply the AI service and retains its return $T-K$ even when no professional report is billed.

To derive the main text's aggregate functions, index case types by complexity $z\in[0,1]$. Each type is a continuum of cases. Write $\bar h(z)$ for reviews per case in existing care, $\widehat h(q,z)$ for the type-specific patient optimum for reviews, and $h^P(a,q)$ for the optimal \emph{common} requirement over adopted cases. Each review pays $p$ and costs the provider $w$, giving margin $\theta=p-w>0$.

With AI, let $T(q,z)$ be the payment for the other services in a case and $K(q,z)$ their provider cost, both measured relative to existing care, so the provider's savings are $B=T-K$. Let $\mathcal H(q,z,h)$ be the monetary value of the change in patient outcomes and access, net of resource costs borne outside the provider. Patient value per adopted case of type $z$ is
$\omega=\mathcal H-T-p(h-\bar h)$ and the provider's gain per case is
$\pi=T-K+(p-w)(h-\bar h)$. Hence
\begin{equation}
 \omega+\pi=\mathcal H-K-w(h-\bar h).
 \label{eq:sum}
\end{equation}
Writing $x=\bar h-\widehat h$ and $s=B-\theta x$ gives $\pi=s+\theta(h-\widehat h)$, the type-level counterpart of the provider term in \eqref{eq:J}. Fixed costs and patient benefits of assisted tasks are included in $K$ and $\mathcal H$, and the corresponding payment is included in $T$.

Within a type, reviews go first to the cases in which they are most valuable, so the marginal health value of additional reviews declines as the review share $h$ rises. Suppose $\mathcal H$ is twice continuously differentiable and $\mathcal H_{hh}=-c<0$ on the relevant review range. The interior patient optimum solves $\mathcal H_h(q,z,\widehat h)=p$. Completing the square around this point yields exactly
\begin{equation}
 \omega(q,z,h)=u(q,z)-\frac c2[h-\widehat h(q,z)]^2,
 \label{eq:appquadratic}
\end{equation}
where $u(q,z)=\omega(q,z,\widehat h)$. Comparative statics in $\theta$ vary $w$ at fixed $p$, which leaves $\widehat h$, $B$, $x$, and patient value unchanged.

\subsection{From a complexity cutoff to adoption}\label{app:aggregate}

Let $F$ be the case distribution, with continuous density $f>0$. Adoption $a=F(\alpha)$ covers cases with complexity $z\leq\alpha$, so $\alpha=F^{-1}(a)$. With a common requirement, the patient optimum is the average type-level optimum among adopted cases,
\begin{equation}
 h^P(a,q)=\frac{1}{a}\int_0^{F^{-1}(a)}\widehat h(q,z)f(z)\dd z.\label{eq:muQ}
\end{equation}
For $a>0$, write $t=h-h^P(a,q)$. Because $h^P$ is this average, $\int_0^{F^{-1}(a)}(h-\widehat h)f\dd z=a t$ and $\int_0^{F^{-1}(a)}(h-\widehat h)^2f\dd z=Q+a t^2$, where $Q=\int_0^{F^{-1}(a)}(\widehat h-h^P)^2f\dd z$ is the dispersion of review needs among adopted cases. Integrating the type-level patient value and provider gain, with $s(q,z)=s(q)$, therefore gives exactly \eqref{eq:J}, with
\begin{equation}
 U(a,q)=\int_0^{F^{-1}(a)}u(q,z)f(z)\dd z-\frac c2 Q.\label{eq:aggregateU}
\end{equation}
Deviations of type-level review needs from $h^P$ cancel in the provider term and enter patient value only through the mismatch cost $cQ/2$. The marginal value $U_a$ equals the value of the marginal type net of its mismatch cost, so the main text's $U_{aa}<0$ means that this value falls along the complexity order. When displaced reviews $x(q)$ are common to all cases, $\widehat h(q,z)=\bar h(z)-x(q)$, so $h^P(a,q)$ is the average requirement in existing care less $x(q)$. The normal example takes the same distribution of reporting difficulty within each type, so $B$ and $x$ are expectations and $s(q,z)=s(q)$. If provider gains instead vary by type, the separate-decision condition is $\lambda U_a+s(q,F^{-1}(a))=0$, and Section~\ref{app:heterogeneity} gives a condition for the gap to keep widening.

\section{Task mechanism and foundations}\label{app:tasks}

\subsection{Proof of Proposition~\ref{prop:task}}\label{app:normalproof}

Under Assumption~\ref{ass:tasks}, write $\mathcal R=B'/x'$, so that $s'(q)=x'(q)[\mathcal R(q)-\theta]$ with $x'>0$. Since $\mathcal R$ is continuous and strictly decreasing, with $\lim_{q\to-\infty}\mathcal R>\theta>\lim_{q\to\infty}\mathcal R$, there is a unique $\widehat q$ with $\mathcal R(\widehat q)=\theta$, and the provider gain strictly increases before $\widehat q$ and strictly decreases thereafter. Since $B(-\infty)=x(-\infty)=0$, $\lim_{q\to-\infty}s(q)=0$, so the maximum is positive, and $\lim_{q\to\infty}s(q)=B(\infty)-\theta x(\infty)$. On the decreasing branch every finite value is strictly above this limit. A negative limit therefore gives exactly one zero $q_0>\widehat q$, followed by a negative provider gain. A nonnegative limit gives a positive provider gain at every finite $q$, with convergence to zero from above at equality.

More generally, let $d\geq0$ be a constant added to the provider gain. The derivative and turning point of $s+d$ are unchanged, its maximum is positive, and its limit is $d-[\theta x(\infty)-B(\infty)]$. A finite zero $q(d)$ therefore exists if and only if $\theta x(\infty)-B(\infty)>d$. At equality there is no finite crossing. Where it exists, $q(d)>\widehat q$, $s(q)+d<0\iff q>q(d)$, and
\[
 \frac{\dd q(d)}{\dd d}=-\frac{1}{s'(q(d))}>0.
\]
Taking $d=\theta t^J/2$, with excess reviews $t^J=\theta/(\lambda c)$ under joint decisions from Proposition~\ref{prop:policy}, proves the threshold $\widetilde q_0=q(\theta t^J/2)>q_0$ under joint decisions and its existence condition $\theta x(\infty)-B(\infty)>\theta t^J/2$.

\subsection{The paired normal example and proof of Corollary~\ref{cor:statics}}\label{app:paired}

In the paired example, $B=b\Phi(q)$ and $x=e\Phi(q-\delta)$. With $\phi=\Phi'$,
\begin{equation}
 \mathcal R(q)=\frac{b}{e}\,\frac{\phi(q)}{\phi(q-\delta)}=\frac be\exp\left(\frac{\delta^2}{2}-\delta q\right),\label{eq:normalratio}
\end{equation}
which is continuous and strictly decreasing from infinity to zero. Assumption~\ref{ass:tasks} therefore holds for all positive $b,e,\theta,\delta$, with $B(\infty)=b$, $x(\infty)=e$, and $\widehat q=\delta/2+\log[b/(\theta e)]/\delta$. The opposition condition of Proposition~\ref{prop:task} becomes $\theta e>b$, and the condition under joint decisions becomes $\theta e-b>\theta t^J/2$.

\emph{Proof of Corollary~\ref{cor:statics}.} Let $\psi$ be any of $b,e,\theta,\delta$. Since $s(q_0)=0$ and $s'(q_0)<0$, implicit differentiation gives $\dd q_0/\dd\psi=-s_\psi(q_0)/s'(q_0)$, which has the sign of $s_\psi(q_0)$. From \eqref{eq:paired},
\[
 s_b=\Phi(q)>0,\qquad s_\delta=\theta e\phi(q-\delta)>0,\qquad
 s_e=-\theta\Phi(q-\delta)<0,\qquad s_\theta=-e\Phi(q-\delta)<0.
\]
Hence $q_0$ rises with $b$ and $\delta$ and falls with $e$ and $\theta$. The same argument applies to $q(d)$ for fixed $d$, holding the other primitives fixed. Since the adoption gap under separate decisions is $-s(q)/\lambda$ by \eqref{eq:separate} and $s_e<0$, the gap is larger for AI that displaces more reviews.

\subsection{Beyond the normal example}

Assumption~\ref{ass:tasks} does not depend on the normal example. Let tasks become feasible at capability requirement $\tau$, with saving $v(\tau)>0$ and displaced reviews $\eta(\tau)>0$, and let $f_B$ and $f_X$ be the densities of assisted and autonomous tasks over $\tau$. Then $B'=vf_B$, $x'=\eta f_X$, and $\mathcal R=(v/\eta)(f_B/f_X)$, which is strictly decreasing if $v/\eta$ and $f_B/f_X$ are non-increasing and at least one strictly decreases, that is, if later tasks bring less saving per displaced review or assisted tasks are concentrated at lower requirements. Neither channel requires every assisted task to precede every autonomous one. The autonomy requirements here are those needed to replace a report, rather than to automate an individual component. For the paired reports of the main text with a general difficulty density $G'$, $\mathcal R=(b/e)\,G'(q)/G'(q-\delta)$, which decreases for every lag $\delta$ if $G'$ is strictly log-concave. Assumption~\ref{ass:tasks} also needs the unscaled density ratio $G'(q)/G'(q-\delta)$ to straddle $\theta e/b$. The normal density satisfies this for all positive $b,e,\theta,\delta$. Other strictly log-concave densities, such as the logistic, may require parameter restrictions. Assistance preceding autonomy alone guarantees only $G(q)\geq G(q-\delta)$.

\subsection{\texorpdfstring{Bundled component tasks}{Bundled component tasks}}\label{app:bundle}

Suppose each report requires $n\geq1$ component tasks, with independent standard normal difficulties $\xi_1,\ldots,\xi_n$ and the same distribution within each case type. AI assists with component $i$ once $q\geq\xi_i$, saving $b/n$, and performs it autonomously once $q\geq\xi_i+\delta$. Savings persist after autonomy, which adds no saving beyond assistance. Let $e>0$ be the expected number of paid reviews per case that AI can displace. This occurs only when all $n$ components are autonomous. Linearity of expectation gives savings $b\Phi(q)$, while independence gives report-replacement probability $\Phi(q-\delta)^n$. Hence
\begin{equation}
 B(q)=b\Phi(q),\qquad x(q)=e\Phi(q-\delta)^n,\qquad
 \frac{B'(q)}{x'(q)}=\frac{b}{en}\,
 \frac{\phi(q)}{\phi(q-\delta)\Phi(q-\delta)^{n-1}}.\label{eq:bundle}
\end{equation}
Writing $\mathcal R=B'/x'$, differentiation yields
\begin{equation}
 \frac{\dd}{\dd q}\log\mathcal R(q)
 =-\delta-(n-1)\frac{\phi(q-\delta)}{\Phi(q-\delta)}<0.\label{eq:bundleratio}
\end{equation}
The ratio falls from infinity to zero: it is the ratio in \eqref{eq:normalratio}, divided by $n\Phi(q-\delta)^{n-1}$. Also, $B(\infty)=b$ and $x(\infty)=e$. Thus Assumption~\ref{ass:tasks} holds, Proposition~\ref{prop:task} applies, and eventual opposition again occurs if and only if $\theta e>b$. The case $n=1$ is exactly \eqref{eq:paired}; with $n>1$, component assistance can lower costs before the complete report can be replaced. Independence and equal savings per component are assumptions of this example. The thresholds generally differ with $n$; Figure~\ref{fig:rollout} retains $n=1$.

\section{Political equilibrium and policy results}\label{app:proofs}

\subsection{Menu-auction equilibrium}

Let $y=(a,h)$. Write $W(a,h;q)=U(a,q)-\frac{c}{2}a t^2$ for patient value and $\Pi(a,h;q)=a[s(q)+\theta t]$ for total provider gain, so that $J=\Pi+\lambda W$ in \eqref{eq:J}. The policymaker maximises contributions $P(y)\geq0$ plus $\lambda W(y)$ and can always choose the maximiser $y^0$ of $W$. An implemented policy $y$ therefore requires $P(y)\geq\lambda[W(y^0)-W(y)]$, so the provider group's net gain $\Pi(y)-P(y)$ is at most $J(y)-\lambda W(y^0)$, which is maximised at the maximiser $y^*$ of $J$. The truthful schedule $P(y)=\max\{0,\Pi(y)-\kappa\}$ with $\kappa=J(y^*)-\lambda W(y^0)$ attains this bound, because it makes $y^*$ a best response for the policymaker, which breaks ties in favour of the provider group \citep{bernheimMenuAuctions1986,grossmanProtectionSale1994}. The same argument applies when policy is restricted to separate decisions, $h=h^P(a,q)$.

\subsection{Proof of Corollary~\ref{cor:rollout} and Proposition~\ref{prop:policy}}\label{app:policyproof}

For the interior results, take $0<a<1$ with slack review bounds and review optima in the range where patient value is quadratic. Write $t=h-h^P(a,q)$. The aggregate objective is
\begin{equation}
 J=\lambda U(a,q)+a s(q)+a\left(\theta t-\frac{\lambda c}{2}t^2\right).\label{eq:political}
\end{equation}
Under separate decisions $t=0$, and strict concavity in adoption gives
$\lambda U_a(a^S,q)+s(q)=0$. Since $U_a$ decreases in $a$ and is zero at $a^P$, $a^S<a^P$ if and only if $s(q)<0$. By Proposition~\ref{prop:task}, this holds if and only if $\theta x(\infty)>B(\infty)$ and $q>q_0$. Differentiating the identity $U_a(a^S(q),q)=-s(q)/\lambda$ along equilibrium gives $-s'(q)/\lambda$, which is positive for $q>\widehat q$ and proves the final statement of Corollary~\ref{cor:rollout}.

At any $a>0$, strict concavity in $t$ yields excess reviews $t^J=h^J-h^P=\theta/(\lambda c)$. Inserting $t^J$ gives the reduced objective
\begin{equation}
 \widetilde J(a,q)=\lambda U(a,q)+a\left[s(q)+\frac{\theta t^J}{2}\right],
 \qquad \frac{\theta t^J}{2}=\frac{\theta^2}{2\lambda c}.\label{eq:reduced}
\end{equation}
Its first-order condition is \eqref{eq:joint}. At $a^S$ its derivative is $\theta t^J/2>0$, so $a^J>a^S$. Comparison with $U_a(a^P,q)=0$ gives $a^J<a^P$ iff $s+\theta t^J/2<0$. Section~\ref{app:normalproof}, with $d=\theta t^J/2$, shows that this holds if and only if $\theta x(\infty)-B(\infty)>\theta t^J/2$ and $q>\widetilde q_0$, where $\widetilde q_0=q(\theta t^J/2)>q(0)=q_0$.

Because the change of variable $t=h-h^P(a,q)$ absorbs the dependence of the common requirement on adoption, optimisation over $h$ is equivalent to optimisation over $t$ and adds no term to these first-order conditions when the review bounds are slack.

Only strict concavity in $t$ is needed for the signs. If excess reviews cost patients $a\,m(t)$, with $m$ strictly convex and $m(0)=m'(0)=0$, then $t^J>0$ solves $\theta=\lambda m'(t^J)$, and the reduced objective adds $a[\theta t^J-\lambda m(t^J)]>0$. Hence $a^J>a^S$, and Proposition~\ref{prop:policy} holds with $\theta t^J-\lambda m(t^J)$ in place of $\theta t^J/2$. The quadratic case $m(t)=ct^2/2$ gives the closed forms.

With the one-report cap and $0\leq h^P(a,q)\leq1$, the conditional review choice at $a>0$ is
\begin{equation}
 t^J(a,q)=\min\left\{\frac{\theta}{\lambda c},1-h^P(a,q)\right\},\qquad
 d(a,q)=\theta t^J(a,q)-\frac{\lambda c}{2}[t^J(a,q)]^2.\label{eq:capgeneral}
\end{equation}
The reduced objective is $\lambda U+a(s+d)$. At an interior adoption choice its first-order condition is
\begin{equation}
 \lambda U_a(a,q)+s(q)+d(a,q)+a d_a(a,q)=0.\label{eq:capadoption}
\end{equation}
Where the cap binds, $d_a=-(\theta-\lambda c t^J)h^P_a$; where it is slack, $d_a=0$. Thus a binding cap and changing case composition need not preserve $a^J>a^S$. If $h^P$ is independent of $a$, as in Figure~\ref{fig:rollout}, $d=d(q)\geq0$ and strict concavity gives $a^J\geq a^S$, strictly when $d>0$ and adoption is interior. Joint under-adoption then occurs exactly when $s(q)+d(q)<0$.

In the normal example with $e=1$ and $h^P=1-\Phi(q-\delta)$, let $\bar t=\min\{\theta/(\lambda c),1\}$. As $q\to\infty$,
\begin{equation}
 s(q)+d(q)\longrightarrow b-\theta+\theta\bar t-\frac{\lambda c}{2}\bar t^2.\label{eq:caplimit}
\end{equation}
If $\theta/(\lambda c)>1$, the cap binds at every capability and this limit is $b-\lambda c/2$. The interior threshold condition cannot be used in this case. Figure~\ref{fig:rollout} has $\theta/(\lambda c)=1/6$, and the cap is slack at both under-adoption thresholds.

\subsection{Patient losses and total surplus}\label{app:welfare}

At $h^J$, the patient-value term of \eqref{eq:J} gives $W(a^J,h^J;q)=U(a^J,q)-a^Jc(t^J)^2/2$. Subtracting this from $U(a^P,q)$ proves \eqref{eq:loss}. The adoption component equals
\[
 U(a^P,q)-U(a^J,q)=\int_{a^J}^{a^P}U_a(v,q)\dd v.
\]
Strict concavity makes it positive whenever $a^J\ne a^P$, including over-adoption. Under separate decisions the loss is $L^S=U(a^P,q)-U(a^S,q)$.

At fixed $a>0$, total surplus $\mathcal S=W+\Pi$ changes with $t$ by $a(\theta t-ct^2/2)$. The total-surplus (efficient) optimum is
\begin{equation}
 h^E=h^P+\frac{\theta}{c},\qquad
 h^J-h^E=\frac{\theta}{c}\left(\frac1\lambda-1\right).\label{eq:social}
\end{equation}
Thus, at interior review policies, reviews under joint decisions exceed the total-surplus optimum at the same adoption exactly when $\lambda<1$.

\section{Quadratic example and figure}\label{app:nesting}

\subsection{Adoption formulas}

Let $U(a,q)=V(q)a-ka^2/2$, $k>0$. At interior policies,
\begin{align}
 a^P&=\frac Vk,&a^S&=\frac{V+s/\lambda}{k},&
 a^P-a^S&=-\frac{s}{\lambda k},\label{eq:oldgap}\\
 a^J&=\frac{V+(s+\theta t^J/2)/\lambda}{k},&
 a^P-a^J&=-\frac{s+\theta t^J/2}{\lambda k},&
 h^J&=h^P+t^J.\label{eq:newhomogeneous}
\end{align}
More generally, differentiating an interior first-order condition, at $a^S$ or at $a^J$, gives
\begin{equation}
 a_q=-\frac{\lambda U_{aq}(a,q)+s'(q)}{\lambda U_{aa}(a,q)},\label{eq:aq}
\end{equation}
because $\theta t^J/2$ does not vary with $q$. Since $U_{aa}<0$, adoption falls iff $-s'>\lambda U_{aq}$. In the quadratic example $U_{aq}=V'$, so both decisions have $a_q=(V'+s'/\lambda)/k$ and the same turning points.

\subsection{Comparing separate and joint decisions}\label{app:regimes}

Write $D=\theta t^J/(2\lambda k)=\theta^2/(2\lambda^2ck)$, so $a^J=a^S+D$. The cost of excess reviews per adopted case is $kD$. Let $L^S=U(a^P,q)-U(a^S,q)$ be the patient loss under separate decisions.

\begin{corollary}[Separate or joint decisions]\label{cor:regime}
At interior policies, $L^J<L^S$ if and only if
\begin{equation}
 a^P-2a^S>\frac{3\theta t^J}{4\lambda k}.\label{eq:regime}
\end{equation}
In particular, $L^S<L^J$ whenever $a^S\geq a^P/2$.
\end{corollary}

\begin{proof}
The adoption formulas are \eqref{eq:oldgap} and \eqref{eq:newhomogeneous}. With all adoption choices interior, the two losses are
\[
 L^S=\frac{k}{2}(a^P-a^S)^2,\qquad
 L^J=\frac{k}{2}(a^P-a^S-D)^2+kD\,a^J.
\]
Their difference is
\begin{equation}
 L^S-L^J=kD\left[a^P-2a^S-\frac{3D}{2}\right],\label{eq:regimeproof}
\end{equation}
which proves the equivalence. If $a^S\geq a^P/2$, the left side of \eqref{eq:regime} is nonpositive, so the inequality fails.
\end{proof}

Joint decisions reduce patient losses only when adoption under separate decisions is below half the patient optimum, and by enough to compensate for the cost of excess reviews.

\subsection{Figure parameters and numerical values}\label{app:figure}

For the illustration in Figure~\ref{fig:rollout}, take the paired reports of \eqref{eq:paired}, $V(q)=V_0+V_1\Phi(\gamma q)$, existing care $\bar h=e$, and
\[
\begin{gathered}
 p=1,\quad w=\tfrac12,\quad\theta=\tfrac12,\quad
 \lambda=\tfrac34,\quad c=4,\quad k=2,\\
 e=1,\quad b=\tfrac14,\quad \delta=1,\quad
 V_0=0.55,\quad V_1=0.70,\quad \gamma=\tfrac14.
\end{gathered}
\]
These values imply the unconstrained excess review $t^*=\theta/(\lambda c)=1/6$ and $h^P=1-\Phi(q-1)$. With the cap $h\leq1$,
\begin{equation}
 t^J(q)=\min\{1/6,\Phi(q-1)\},\qquad
 d(q)=\theta t^J(q)-\frac{\lambda c}{2}[t^J(q)]^2,\qquad
 a^J(q)=\frac{V(q)+[s(q)+d(q)]/\lambda}{k}.
 \label{eq:figurecap}
\end{equation}
The cap binds for $q<q_c\equiv1+\Phi^{-1}(1/6)\simeq0.033$, where $h^J=1$ and $s+d=b\Phi(q)-\lambda c\Phi(q-1)^2/2$. For $q\geq q_c$, $d=1/24$ and $a^J-a^S=1/36$, as in the interior formulas above. The joint patient loss is $[s+d]^2/(2\lambda^2k)+a^Jc(t^J)^2/2$. The figure is produced by the Python script \texttt{figures/make\_figure1\_revision.py}, provided with this Supplementary Material.

Outside the capped range, adoption falls between the zeros of $\lambda V'+s'$, at $q_1\simeq0.238$ and $q_2\simeq2.742$, and rises elsewhere. The thresholds are $q_0\simeq0.686$ and $\widetilde q_0\simeq1.014$. The shortfall $a^P-a^J$ keeps widening after adoption recovers and tends to $(\theta e-b-\theta t^J/2)/(\lambda k)=5/36$, because $s$ decreases for $q>\widehat q$ and is bounded. For $q<q_c<q_0$, $s>0$, so $a^P<a^S<a^J$ and separate decisions lose less. The script checks numerically that all three adoption policies are interior, that joint adoption rises for $q<q_c$, and that $L^S<L^J$ at every plotted $q$.

\section{Extensions}\label{app:extensions}

\subsection{Direct payments for AI use}\label{app:direct}

Payers pay for services listed in their fee schedules, so direct compensation for displaced reviews takes the form of a payment per AI use, such as Medicare's payment for the autonomous retinal examination \citep{abramoffReimbursementFramework2022}. Each dollar of such a payment costs patients a dollar, collectively through premiums and taxes. Suppose the policymaker can also pay providers $\rho\geq0$ per adopted case, independently of reviews. The payment lowers $W$ and raises $\Pi$ by $a\rho$, adding $a(1-\lambda)\rho$ to \eqref{eq:political}, so the review rule at fixed adoption is unchanged. The equilibrium requirement can change if the payment changes adoption and hence $h^P(a,q)$. The chosen payment is zero if $\lambda>1$, and if $\lambda<1$ it must be capped by a budget or coding constraint.

At the main model's interior joint policy, excess reviews raise provider income by $\theta^2/(\lambda c)$ per adopted case at a patient cost of $\theta^2/(2\lambda^2c)$. A direct payment of equal provider value costs patients $\theta^2/(\lambda c)$. At this interior policy, reviews are therefore the cheaper transfer iff $\lambda>1/2$. With the cap $h\leq1$, the chosen excess review is $t=\min\{\theta/(\lambda c),1-h^P\}$, and its patient cost per dollar of provider gain is $ct/(2\theta)\leq1/(2\lambda)$. Reviews are then cheaper than an equal direct payment iff $t<2\theta/c$, so $\lambda>1/2$ is sufficient but not necessary. Their clinical value offsets part of their fee, since the first excess reviews at the patient optimum have zero net cost to first order. This comparison holds adoption and the payment system fixed. Compensation through reviews can also reflect lower visibility \citep{coateMorrisTransfers1995} or interest groups' preference for indirect instruments \citep{grossmanProtectionSale1994}, despite pressure toward efficient transfers \citep{beckerPressureGroups1983}.

\subsection{Who retains the return from AI-only care}\label{app:ownership}

The main model assumes that the organised provider group keeps the saving $b$ even after AI replaces its paid report. Ownership of complementary services can change which providers capture such returns \citep{chenGertlerYangOwnership2016}. To isolate this incidence, keep patient outcomes, patient-facing payments, $U$, and the technology fixed, and let the organised group receive only $r\in[0,b]$ of the return on a case that AI handles alone, with $b-r$ going to an unorganised provider. This is not a change in the payer's fee. Take the paired normal example with one existing report per case ($e=1$), so $h^P=1-\Phi(q-\delta)$.

An assisted case with a paid report gives the group $b$ relative to existing care, and an AI-only case gives $r-\theta$. An excess paid report turns an AI-only case back into assisted care, so it is worth $\alpha_r\equiv\theta+b-r$ to the group rather than $\theta$. The provider gain at the patient optimum and the joint review policy are
\begin{equation}
 s_r(q)=b\Phi(q)-\alpha_r\Phi(q-\delta),\qquad
 t^J_r(q)=\min\left\{\frac{\alpha_r}{\lambda c},\Phi(q-\delta)\right\},\label{eq:ownershipgain}
\end{equation}
where the upper bound is the cap $h\leq1$. For $r=b$ this is the main model. Since $s_r'(q)=\phi(q-\delta)[b\exp(\delta^2/2-\delta q)-\alpha_r]$ changes sign once and $s_r(\infty)=r-\theta$, support turns to opposition at a finite $q_0(r)$ if and only if $r<\theta$. Excess reviews add $d_r=\alpha_rt^J_r-\lambda c(t^J_r)^2/2>0$ to the return on adoption, so joint decisions still increase adoption at every capability, and joint under-adoption, where it occurs, starts later than $q_0(r)$.

At the Figure~\ref{fig:rollout} parameters, moving from $r=b=1/4$ to $r=0$ moves the separate-decision opposition threshold from $0.686$ to $0.068$ and the joint threshold from $1.014$ to $0.684$. The cap then binds for $q<0.326$. The larger review incentive does not fully offset the earlier loss of AI-only income. With the same patient-value primitives, joint decisions yield lower patient loss only for approximately $2.85<q<4.12$. Thus the main figure's patient-loss ranking is an illustration, not a general consequence of separating decisions. If $r$ instead changes total reimbursement or resource costs, $U$ must also change and these formulas do not apply.

\subsection{Nonpositive margins}

The main results assume $\theta>0$. With no marginal review income, $\theta=0$, the provider gain is $B\geq0$ and there are no excess reviews. Under a fixed payment per case, the incremental fee for an additional review is zero, so $\theta=-w<0$. Displacing reviews then saves the provider resources, and $s=B+w x\geq0$.

If the quadratic patient value remains valid over the relevant range, the constrained requirement under joint decisions at positive adoption is
\[
 h^J(a,q)=\max\{0,h^P(a,q)+\theta/(\lambda c)\}.
\]
Thus influence favours fewer reviews. The cap $h\leq1$ in Figure~\ref{fig:rollout} does not alter this nonpositive-margin solution when $h^P\leq1$.

\subsection{\texorpdfstring{Case-specific provider gains}{Case-specific provider gains}}\label{app:heterogeneity}

Retain adoption by a complexity cutoff and take uniformly distributed case types $z\in[0,1]$, so adopted types satisfy $z\leq a$. Let $s(q,z)=s_0(q-\kappa z)$, with $\kappa>0$ and $s_0$ the main model's provider gain: more complex cases require greater AI capability. Take $U(a,q)=V(q)a-ka^2/2$, $k>0$, with differentiable $V$. Under separate decisions the policymaker maximises $\lambda U(a,q)+\int_0^a s_0(q-\kappa z)\dd z$. At interior patient-optimal and political adoption, write $u=q-\kappa a^S(q)$. The first-order and strict second-order conditions are
\begin{equation}
 \lambda[V(q)-ka^S]+s_0(u)=0,\qquad
 D\equiv\lambda k+\kappa s_0'(u)>0.\label{eq:heterofoc}
\end{equation}
Along this interior solution, implicit differentiation gives $\dd a^S/\dd q=[\lambda V'(q)+s_0'(u)]/D$. Thus the marginal adoption gap $G(q)=U_a(a^S(q),q)=V(q)-ka^S(q)$ satisfies
\begin{equation}
 G'(q)=\frac{-s_0'(u)[k-\kappa V'(q)]}{\lambda k+\kappa s_0'(u)}.\label{eq:heterogap}
\end{equation}
Equivalently, since $\lambda G(q)=-s_0(u)$ by \eqref{eq:heterofoc}, $G'(q)=-s_0'(u)\,u'(q)/\lambda$, where $u'(q)=\lambda[k-\kappa V'(q)]/D$ is the change in the effective capability of the marginal case. Under-adoption occurs when $s_0(u)<0$, which also implies $s_0'(u)<0$ by Proposition~\ref{prop:task}. Given $D>0$, the gap then widens exactly when $k>\kappa V'(q)$, is locally unchanged at equality, and narrows when the inequality reverses. Since $a^P=V/k$, the widening condition is $\kappa\,\dd a^P/\dd q<1$. If adoption expands into more complex cases sufficiently quickly, AI can become less autonomous for the marginal case despite increasing capability. This composition effect can reverse the widening result; the common-gain model excludes it.

\end{document}